\documentclass[letterpaper,twocolumn,10pt]{article}
\usepackage{usenix-2020-09}

\usepackage[utf8]{inputenc}

\usepackage{tikz}
\usepackage{amsmath}
\usepackage{titlesec}
\usepackage{xspace}
\usepackage{enumitem}
\usepackage{subfigure}
\usepackage{mathtools}
\usepackage{xurl}
\usepackage{multirow}
\usepackage{algorithm}
\usepackage{algpseudocode}

\usepackage{titling}

\usepackage{soul}
\usepackage{appendix}
\usepackage{float}
\usepackage{titlesec}

\usepackage{pifont}

\newcommand{\name}{{\sc KnapsackLB}\xspace}
\newcommand{\ie}{\emph{i.e.,}\xspace}
\newcommand{\eg}{\emph{e.g.,}\xspace}

 \newcommand{\ngs}[1]{}

\titlespacing\section{2pt}{2pt plus 1pt minus 1pt}{2pt plus 2pt minus 2pt}
\titlespacing\subsection{2pt}{2pt plus 1pt minus 1pt}{2pt plus 2pt minus 2pt}
\titlespacing\subsubsection{2pt}{1pt plus 1pt minus 1pt}{0pt plus 2pt minus 2pt}

\renewcommand{\baselinestretch}{0.98}

\begin{document}
\sloppy
\setlength{\droptitle}{-3em}   
\title{\name: Enabling Performance-Aware Layer-4 Load Balancing} 


\author{
{
Rohan Gandhi (Microsoft Research), Srinivas Narayana (Rutgers University)
}
}
\date{\vspace{-3ex}}
\maketitle

\section*{Abstract}
    Layer-4 load balancer (LB) is a key building block of online
    services.  In this paper, we empower such LBs to adapt to different and dynamic performance of  backend instances (DIPs). Our system, \name, is
    generic (can work with variety of LBs), does not require
    agents on DIPs, LBs or clients,
    and scales to large numbers of DIPs.
    \name uses judicious active probes to learn a mapping from LB weights to the response latency of each DIP, and then applies
    Integer Linear Programming (ILP) to calculate LB weights that
    optimize latency, using an iterative method to
    scale the computation to large numbers of DIPs.  Using testbed
    experiments and simulations, we show that \name load balances
    traffic as per the performance and cuts average latency by
    up to 45\% compared to existing designs.

\section{Introduction}
\label{sec:intro}

A high performance layer-4 load balancer (L4 LB) is one of the key
building blocks of online services. Individual services scale by running on multiple backend servers or VMs with unique IPs (referred as
{\em DIPs}). The service exposes a small number of virtual IPs, termed {\em
  VIPs}, to receive traffic from outside the service. The LB receives traffic coming on the VIPs and distributes
it across DIPs. Recently, there has been significant interest in
improving the scale, availability and cost of the
LB~\cite{ananta:sigcomm13, maglev:nsdi16, duet:sigcomm14,
  silkroad:sigcomm17, beamer:nsdi18, cheetah:nsdi20, tiara:nsdi22, faild:nsdi18}. 


Ideally, an L4 LB should split the traffic to provide \textit{best performance} (e.g., latency) across the DIPs. To do so, many of the LBs assume that capacity of the DIPs is same or at least known. When the capacity is same, LBs can simply split traffic equally getting uniform performance on all DIPs.  Else, it resorts to algorithms such as weighted round-robin or least-connection to load balance the traffic\cite{ananta:sigcomm13, maglev:nsdi16, cheetah:nsdi20, haproxy:web} where the weights are set by the operators based on the capacities of DIPs. However, there is a growing trend of DIPs exhibiting capacities that change dynamically, especially in virtualized clusters. Prior work has shown that the capacity of VMs with the same type (\eg D3.medium) may vary up to 40\% due to noisy neighbors: VMs on the same host compete for shared
resources such as caches and memory buses, resulting in variation. Recent proposals
and new offerings from cloud providers introduce dynamic changes to the number of vCPUs assigned to VMs (\S\ref{sec:back}). Together, these factors make
capacities variable across DIPs and over time. 
 
In such a setup where the capacity of the DIPs change dynamically, we find that existing LBs fall short in achieving a goal of providing best performance. We consider well-known LBs and
algorithms: (a) HAProxy LB -- one of the widely
used open source LB with all available algorithms, (b) Microsoft Azure
LB. We find that the existing algorithms fall short in splitting traffic conforming to the capacities resulting in poorer performance on some of the DIPs (\S\ref{sec:back:limit1}). Interestingly, the {\em least connections} algorithm, which is specifically built to adapt to variable DIP performance, also fell short. As a result, the requests going to the over-utilized DIPs suffered poor performance (2$\times$ higher
latency than other DIPs). Lastly, many existing LBs often remove DIPs suffering from poor performance, instead of adapting to their performance. 

In this paper, we revisit the first order question on how L4 LBs split traffic. We argue that L4 LB problem should be modeled as a \textit{Knapsack problem}\cite{knapsack:book} to pack the load to optimize service performance while conforming to static/dynamic capacities of the DIPs. The key challenge is determining how much load is safe to direct to or away from a DIP without adversely impacting service performance.

This paper presents \name, a ``meta'' LB design that enables other L4 LBs
to provide best performance. \name does
not intent to add to the impressive list of existing LB designs. Instead, \name
enables setting the DIP weights for any LB that can implement
weighted load balancing. A significant number of LB systems provide an
interface to specify the weights, including AVI\cite{avi:web},
Nginx\cite{nginx:web}, Duet\cite{duet:sigcomm14}, and
HAProxy\cite{haproxy:web}. \name frees admins from configuring weights
according to DIP performance that are either static or dynamic. In
doing so, it leverages the availability and scalability of existing LB
designs while endowing them to provide best performance.


At a high level, using SDN principles, \name decouples weight computation from LBs and does it at a central controller. This enables \name to make optimal decisions with a \textit{global view} across DIPs. We use request-response (service) latency as a proxy for the performance\footnote{We focus on services where latency matters (e.g., web services).}. \name uses active probing
judiciously to learn a {\em per-DIP weight-latency curve}: a mapping
from a weight to the DIP's average response latency, should
that weight be applied to that DIP. This mapping is used to drive an
Integer Linear Program (ILP) that computes LB weights to pack load
into DIPs, while minimizing the average response latency
across DIPs. The \name\ controller uses the LB's existing
interface to configure DIP weights. In doing so, \name does not require any changes to DIPs, clients, or the LB.

\name addresses three key technical challenges to achieve its goals
(\S\ref{sec:overview:key}). First, issuing probes naively for the
set of all possible weights would require an impractically high number
of latency measurements across hundreds of DIPs. \name
implements just a few measurements (\S\ref{sec:design:curve},
\S\ref{sec:design:measure}), each adapting from the results of the
past measurements, and runs curve-fitting to learn a weight-latency
curve that works well across a wide range of weights.

Second, running an ILP that handles hundreds of DIPs with fine-grained
DIP weight settings is computationally expensive. \name uses a
multi-step ILP to substantially speed up the calculation of DIP
weights. Specifically, in each step, \name runs an ILP that
progressively zooms into finer-grained weight settings per DIP,
instead of calculating weights in one shot
(\S\ref{sec:design:runtime}).

Third, a weight-latency mapping learned at some fixed load
arriving at an LB could become stale. Specifically, the
load placed on a DIP, and hence its response latency, depends
on both the LB's aggregate incoming load and the DIP's
weight. \name avoids learning a new weight-latency curve for each
aggregate load. Instead, \name adapts quickly to variations in traffic
rate by rescaling and shifting its learned weight-latency curve
(\S\ref{sec:dynamics}).

We evaluate a 41-VM prototype implementation of \name (\S\ref{sec:impl}) on  Azure and also at larger scale using simulations. Our results (\S\ref{sec:testbed}) show: (a) \name can
adjust the weights as per the performance of the DIPs. (b) In doing so,
it cuts the latency up to 45\% for up to 79\% requests. (c) \name
adapts to changes in the cluster in terms of capacity, traffic,
and failures. (d) \name solves ILP quickly using optimizations. (e) \name can work with wide range of other LBs. (f) \name incurs only a
small overhead in terms of CPU cores and costs.

This work does not raise any ethical issues.

\section{Background}
\label{sec:back}

\begin{figure}[t]
\centering
\includegraphics[width=0.37\textwidth, page=1]{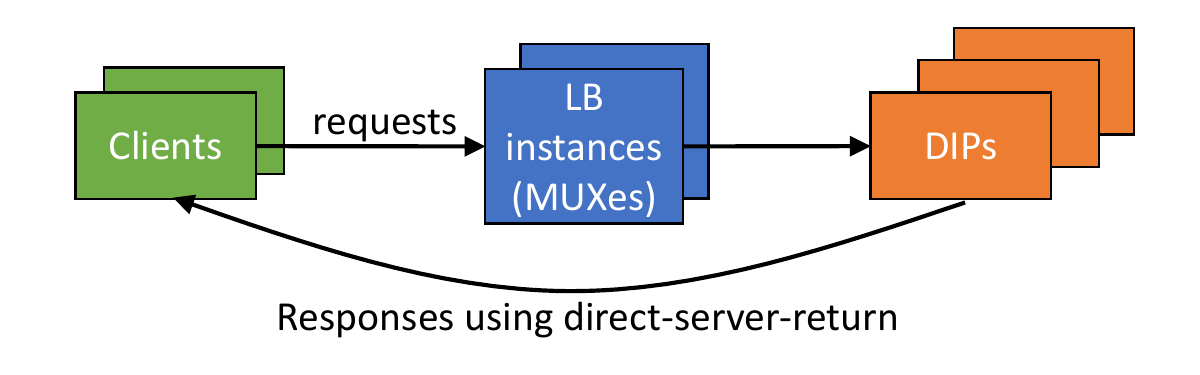}
\caption{LBs run on multiple instanced called MUXes.}
\vspace{-0.1in}
\label{fig:back:lb}
\end{figure}

An online service deploys multiple servers (called DIPs) to scale and
to provide high availability. The DIPs run behind a load balancer (LB)
that exposes one or more virtual IPs (VIPs) to receive the traffic
from outside the service. In this paper we focus on layer-4 LB that
splits the traffic across DIPs using TCP/IP fields. VIPs offer
multiple benefits for scalability, security and high
availability\cite{ananta:sigcomm13}.

Fig.\ref{fig:back:lb} shows terminologies in using L4 LBs. LB designs run on multiple instances (called MUXes). MUXes intercept the VIP traffic  and split them across DIPs\cite{ananta:sigcomm13, maglev:nsdi16}.

\textbf{DIP selection:} MUXes need to process packets with high throughput and low latency. Thus, they pick DIPs for new \textit{connections} using fast  algorithms such as (weighted) round robin, hash over TCP/IP fields, power of two or (weighted) least connection\cite{lbalgo:web}. However, a long standing assumption for these algorithms to work well is that the \textit{DIPs have same or known capacities} (capacity = max. throughput of a DIP) so that MUXes can try weights to get uniform load balancing to not overload DIPs and have good performance across DIPs. However, there is a growing trend towards DIPs of the same service to have different/dynamic capacities as detailed next. LB algorithms fall short in automatically optimizing performance in such cases as detailed next.


\subsection{Limitation-1: LBs Cannot Adapt to Dynamic Capacities}
\label{sec:back:limit1}

In this section, we describe: (a) some of the reasons on why DIPs can have capacities changing dynamically, (b) how existing LBs fall short in providing best performance when capacities of DIPs change dynamically.

\subsubsection{DIPs Can Have Capacities Changing Dynamically}
\label{sec:back:limit1:reason}
\vspace{0.05in}

Existing and recent works have shown that the VMs in cloud have dramatically
different capacities even when they fall in the same VM type (\eg
D3.medium in Azure) (more details in \S\ref{sec:related}). The change
in capacities mainly stems from {\em noisy neighbors,} \ie the
contention from the VMs on the same host. Even if the vCPUs are
isolated, other resources including CPU
caches\cite{contention:euro18}, main memory\cite{contention:micro15,
  contention:hpca13}, and the memory bus are shared. VMs on the same
host contend for such shared resources dynamically, translating to
dynamic DIP capacities. Additionally, our private conversation with a major cloud provider also indicated \textit{dynamic over-subscription}, \ie the number of vCPUs sharing the same physical cores changes dynamically depending on customer demand, resulting in changes in capacities. Importantly, the \textit{changes in capacity are variable and may occur at any time}. As we show in the next section, such change in capacities impacts the performance and existing LBs do not react well to the dynamic capacities.


\subsubsection{LBs Dont Adapt to Dynamic Changes in DIP Capacities}
\label{sec:back:limit1:limit}
\vspace{0.05in}

\begin{figure*}
\centering
    \begin{minipage}{.3\textwidth}
        \centering
        \includegraphics[width=1\textwidth]{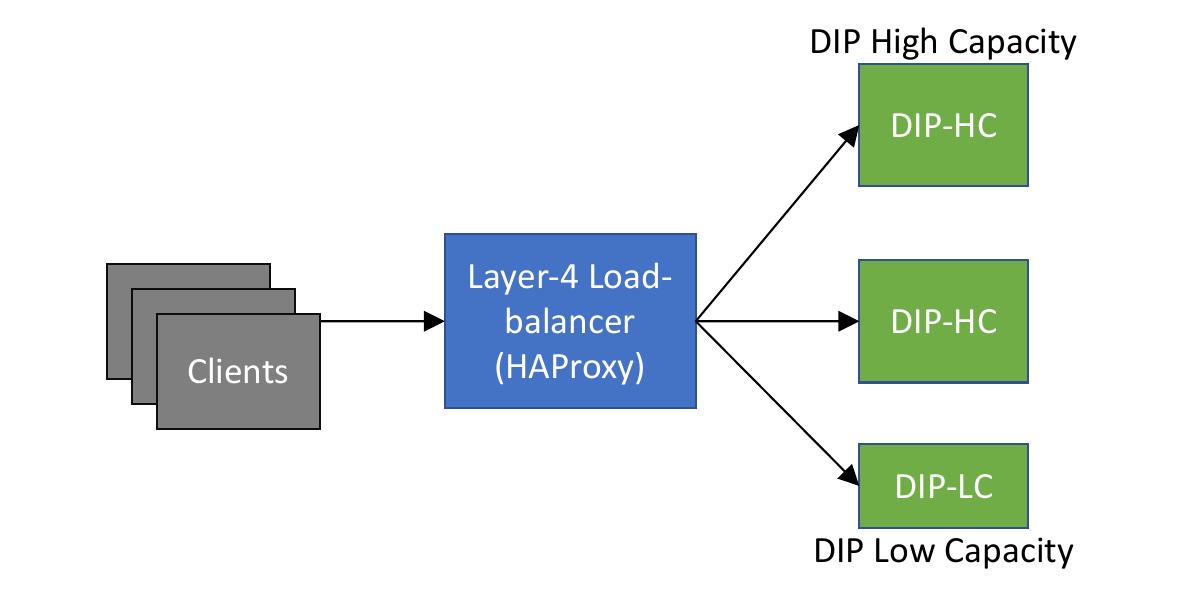}
        \vspace{-0.2in}
        \caption{Setup for evaluating RR and LCA policies in HAProxy LB.}
        \label{fig:back:setup}
    \end{minipage}%
    \hspace{0.2cm}
    \begin{minipage}{0.3\textwidth}
        \centering
        \includegraphics[width=1\textwidth, page=1]{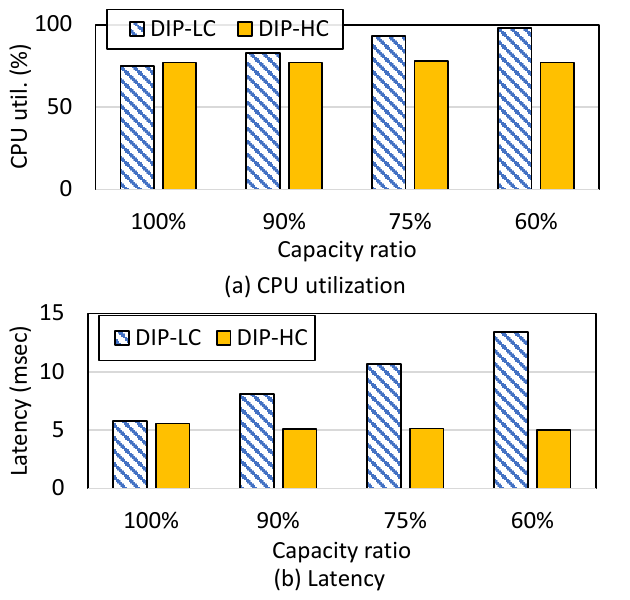}
        \vspace{-0.2in}
        \caption{Performance of RR with changes in capacity.}
        \label{fig:back:limitrr}
    \end{minipage}%
    \hspace{0.2cm}
    \begin{minipage}{0.3\textwidth}
        \centering
        \includegraphics[width=1\textwidth,page=2]{figs/back-limits.pdf}
        \vspace{-0.2in}
        \caption{Performance of LCA  with changes in capacity.}
        \label{fig:back:limitlc}
    \end{minipage}%
\vspace{-0.1in}
\end{figure*}


We consider two L4 LBs: (a)
HAProxy\cite{haproxy:web}, a widely used open-sourced LB, and (b)
Microsoft Azure.

\textbf{HAProxy.} Fig.\ref{fig:back:setup} shows our experimental setup. We
have HAProxy running on a 8-core VM. We have three DIPs -- DIP-HC (x2)
and DIP-LC (HC/LC = High/Low Capacity) on 2-core VMs each. All three
DIPs run web servers that compute cache intensive task for every HTTP request. DIP-HC VMs have the same
capacity; we change the capacity of DIP-LC by running varying number
of copies of an antagonist process that thrashes caches (and partially consumes CPU). All VMs run in Azure on DS series CPU running Ubuntu 20.04.

We evaluate all algorithms in HAProxy and present results for 2
popular algorithms -- (a) round-robin (RR), (b) least-connection
(LCA). RR simply rotates new connections across 3 DIPs. LCA uses the
number of active connections as a proxy for load and selects the DIP with the least number of connections when the new connection appears.

\textbf{Metric of interest:} For both policies, the metric of interest is request-response latency  (for client requests) for all DIPs. We take average for 10K requests. 

Fig.\ref{fig:back:limitrr} shows the CPU utilization and request-response
latency for RR. We start with all DIPs of same capacity. The X-axis
shows the ratio of capacity of DIP-LC to DIP-HC,  denoted as "Capacity
ratio." When capacity ratio is 100\%, the CPU utilization on all DIPs
is close to 80\%. Next, to emulate the dynamic/different capacity, we
deploy the antagonist process. Note that, the impact of noisy neighbors could be variable and unpredictable. We want to evaluate if RR (and LCA later) can adapt to new capacities automatically without manual intervention from the service owners. However, we found that HAProxy continues to split the traffic in the same way as before (equal split). As a result, for lower capacity ratio, as DIP-LC has lower capacity, it hits close to 100\% utilization before DIP-HC VMs. 


The CPU imbalance also translates in difference in request-response latencies (Fig.\ref{fig:back:limitrr}(b)). Because of dynamic reduction in capacities and HAProxy continuing to send traffic at same rate, DIP-LC sees higher CPU utilization and higher latencies for the same incoming traffic rate. The imbalance in latencies across DIPs only gets worse as capacity ratio decreases. 

Now we turn to LCA and evaluate whether it can balance load to provide lower latencies. We keep the same setup. Fig.\ref{fig:back:limitlc}(a) and
Fig.\ref{fig:back:limitlc}(b) show the CPU utilization and end-to-end
latency as we decrease the capacity ratio. Surprisingly, despite LCA
being a policy that considers performance, our observations are similar to
RR. As shown in Fig.\ref{fig:back:limitlc}(a), as we decrease the
capacity ratio, there is imbalance in the CPU utilization on DIP-LC compared to DIP-HC VMs. While the imbalance is slightly smaller than RR, it is still prevalent. LCA also results in CPU fully utilized on DIP-LC whereas DIP-HC VMs are under-utilized resulting in higher latencies on DIP-LC compared to DIP-HC VMs.

At first glance, it may seem that LCA will adapt to different performance by assigning more new connections to the DIP with the
fewest connections at that time. A faster DIP will mostly result in fewer connections than the slower counterparts. However, in doing so, it effectively
\textit{splits the number of concurrent connections equally} among
DIPs. DIPs that finish connections faster will get more connections. However, some DIPs (with lower capacity) can get overwhelmed with many concurrent connections. In such cases, LCA never
reduces the concurrent connections on such DIPs. In the previous example with capacity ratio of 60\%, DIP-LC got roughly 27\% of all
connections; all such connections going to DIP-LC overwhelmed DIP-LC and caused higher latency. 

The other algorithms also exhibited CPU and latency imbalance (not shown due to limited space). 

\begin{table}
\centering 
\caption{Load imbalance using Azure L4 LB.}
\label{tab:back:azurelb}
\begin{tabular}{|c|c|c|}
\hline
DIPs & CPU utilization & Latency\\
\hline
DIP-LC &  84\% & 7.18 msec\\
\hline
DIP-HC & 51\% & 5.00 msec\\
\hline
\end{tabular}
\vspace{-0.1in}
\end{table}

\textbf{Limitations using Azure public L4 LB:} We repeated the above
experiment using Azure L4 LB. We used above 3 DIPs with Azure
LB. Azure LB only provides IP 5-tuple-based load
balancing\cite{azurelb:web}.
Table \ref{tab:back:azurelb} shows the latency and CPU utilization on
DIP-HC and DIP-LC. Here DIP-LC has 60\% capacity compared to
DIP-HC. Unsurprisingly, as IP 5-tuple-based load balancing is not
inherently built to provide best performance, we see imbalance in terms of the CPU
utilization and latency -- latency of DIP-LC to be
43\% higher than DIP-HC.  

These results show that HAProxy and Azure LB fail to provide best performance (lower latencies) by automatically adapting traffic based on performance. 

\subsection{Limitation-2: LBs Cannot Adapt to Differences in Performance}
\label{sec:back:limit2}

\subsubsection{DIPs Can Have Different Performance}
\label{sec:back:limit2:reason}
\vspace{0.05in}
Our private conversation with a major online service owner (using more than 100K VMs) indicated that they don't always use pay-as-you-go model where they dynamically scale up/down VMs based on the demand. The key reason is that public cloud providers don't always have capacity needed, especially during peak hours. As a result, the service owners don't release their VMs. Instead, they reassign the VMs to different parts of their service. Such a reassignment results in DIPs of different types (e.g., DS\cite{dsazure:web} and F\cite{fazure:web} type) for the same VIP, resulting in DIPs with different performance in the same DIP-pool of a service. We observed F-type VMs to provide 15-20\% lower latency for simple request-response traffic compared to DS-type VMs in Azure. 

\subsubsection{LBs Dont Adapt to Differences in Performance}
\label{sec:back:limit2:limit}
\vspace{0.05in}

In this experiment, we show that LBs dont react to static differences in performance. We have two DIPs behind LB -- one each from DS- and F- series in Azure with same number of cores. Both the DIPs run web server. Again, our metric of interest is request-response latency -- whether LBs provide the best overall latencies.  

However, both HAProxy (using RR) and Azure LB \textit{split the traffic equally among such VMs} that did not yield optimal latency. Ideally, LBs should have sent more traffic to F-type VMs to lower overall latency. Next, HAProxy (using LCA) sent 2\% more requests to F-type VMs not leveraging the full potential of F-type VMs. We can further lower the latency by carefully sending more traffic to F-series VM.  

\subsection{Changing Weights to Adapt to Capacity} 
As mentioned previously, one of the key limitation of HAProxy and Azure LBs is that they did not automatically adjust the traffic split to provide best performance. However, manual intervention by changing the weights to split the traffic is also not trivial. If some DIPs are suffering from poor performance (hotspots), if too little traffic is taken away, that may not clear the hotspots. On other extreme, if too much traffic is taken away, it risks hotspots moving somewhere else. \name relieves service operators from dealing with traffic split for optimal performance.


\section{\name}
\label{sec:overview}

\subsection{Goals}

We present \name, a ``meta'' LB  designed to meet the following
goals:

\textbf{Performance-optimized LB}: 
Our vision is that \name should enable other L4 LBs to split traffic to provide the best performance. Admins can plugin any DIP, and leave it to \name to provide best performance \textit{without taking any hints about capacities or performance}. \name should adjust load even when the performance changes dynamically. In essence, \name must free-up admins from the labor of setting up performance-optimized LBs.

\textbf{Generality}: \name intends to work with existing LB
designs \textit{without any changes to MUXes, DIPs or clients}. 





\textbf{Zero-touch and agent-less design:} \name should not run any agents on
DIPs, clients or MUXes. Consequently, \name should work without access to DIP SKU, CPU utilization, or any
performance counters on DIPs. This goal helps customers maintain privacy about their resource usage, and helps \name reduce deployment overheads.

\textbf{Everything online}: \name should not require any offline
profiling data. Using just the IP addresses of the DIPs, \name should
perform all its functions online.


\subsection{\name Overview and Architecture}
\label{sec:overview:overv}

%

\label{sec:overview:arch}

\begin{figure}[t]
\centering
\includegraphics[width=0.45\textwidth, page=3]{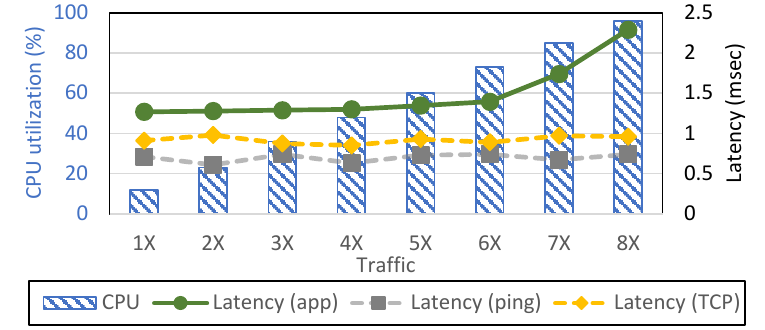}
\caption{Impact of increasing weights (traffic) on latency (y2 axis) and CPU utilization (y1 axis). TCP and ICMP pings are unaffected by changing weights (traffic).}
\label{fig:design:weight}
\end{figure}

\begin{figure}[t]
\centering
\includegraphics[width=0.47\textwidth]{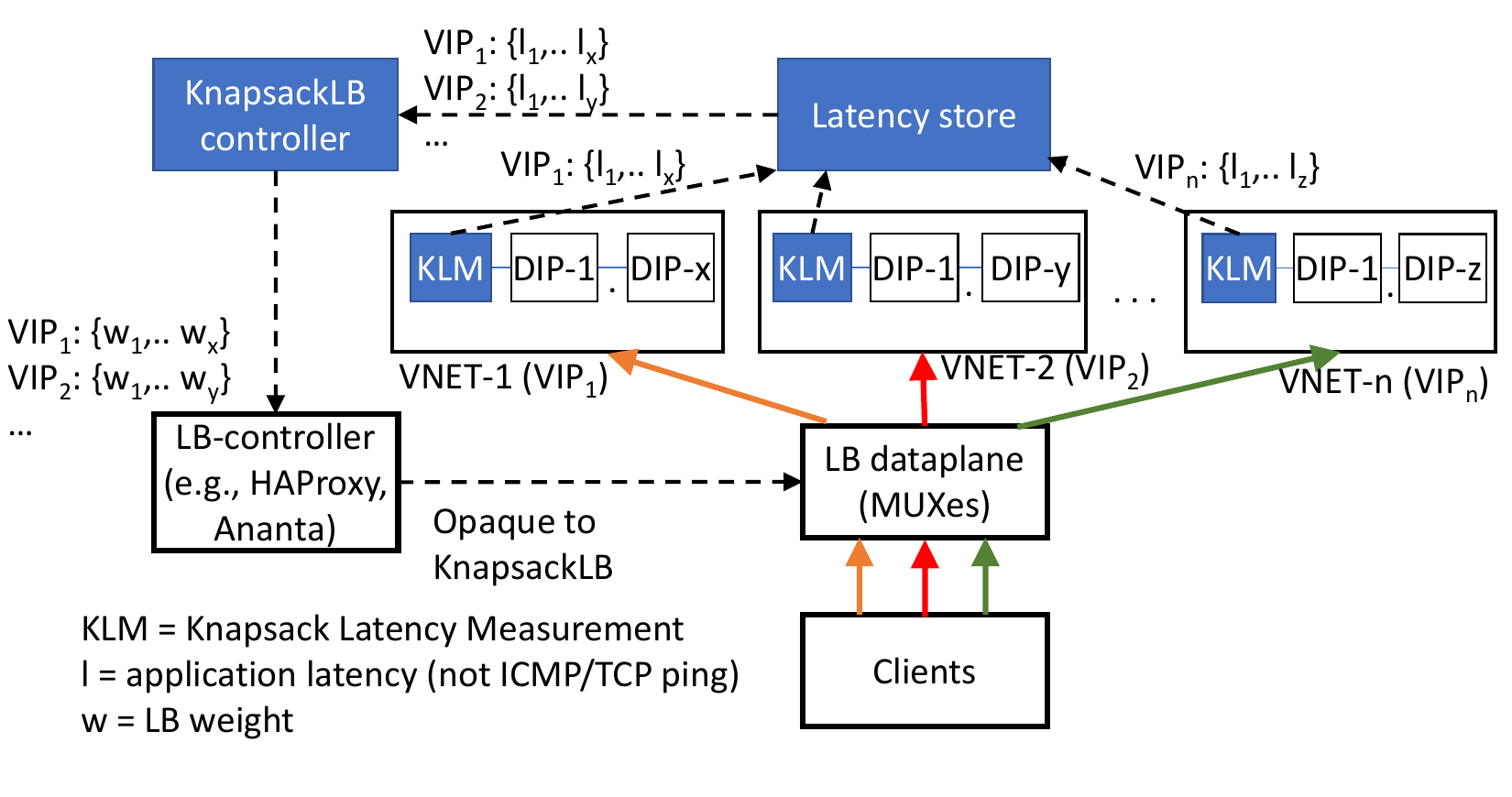}
\caption{\name architecture. Blue boxes denote \name components. Solid and dotted lines denote data traffic and control signals respectively.}
\label{fig:overview:arch}
\end{figure}


\name casts the LB problem as a Knapsack problem to pack the load as per the capacities of the DIPs to optimize the service latency (\S\ref{sec:ilp}). We focus on DIPs running services where latency matters (e.g., web services). As mentioned above \name intends to work with existing LB designs without agents on MUXes, DIPs or clients. 

\name has three key aspects: (a)
gauging \textit{DIP performance vs. weights}, through an active
probing approach (\S\ref{sec:design:curve}, \S\ref{sec:design:measure}) that works without DIP, MUX,
or client modification, (b) weight computation at a centralized
controller, that uses a multi-step ILP (\S\ref{sec:design:runtime}) with the goal of
packing load into DIPs to minimize average response latency, and
(c) programming the DIP weights on the fast data path.

\textbf{Using weights to control traffic:} We observe that many existing LB designs provide an interface to specify the weights for splitting the traffic. \name uses this interface to program the weights. This way, \name can support a
variety of other LBs. Wherever such an interface is not provided, we
use DNS based LB (an example in
\S\ref{sec:eval:other}). Importantly, the MUXes themselves remain unmodified and
their throughput and latency are unaffected by \name.

\textbf{Latency as a proxy for service performance:} \name optimizes for the \textit{application-level latency} using requests that are served by the service running on the DIPs. We use such a latency as a proxy for service performance. As shown in Fig.\ref{fig:design:weight}, ICMP and TCP pings (using SYN/SYN-ACK) do not reflect the load on the DIP as they are handled by the OS. In contrast, application (app) requests are handled by the service on the DIPs and reflect the performance of the service (e.g., queues formed at applications). 

\textbf{Architecture:} As shown in  Fig.\ref{fig:overview:arch}, \name has three
loosely coupled components: (a) KLM (\name Latency Measurement), (b)
latency store, and (c) controller.

KLM resides in each customer-level virtual network (VNET). For
simplicity, we assume there is one externally-visible VIP per
VNET. KLM periodically measures the latency for requests from each
DIP.  KLM directly
measures the latency at the service-level (\eg HTTP requests)
using service URLs provided by the administrators.  KLM also
sends requests directly to the DIPs (using IPs of DIPs; bypassing MUXes) to eliminate
the interference of the MUXes on DIP latency. The latency store persists the measurements from the KLM. Implementing
measurements that directly measure application-layer latencies and storing them separately
obviates the need for any instrumentation on DIPs, MUXes or clients, and enables \name to avoid changes to the MUXes, DIPs and clients,
broadening \name's utility to a variety of LBs and settings. 

The \name controller consumes DIP latencies from the latency store. It
computes the weights using Integer Linear Program (\S\ref{sec:ilp}) and sends
them to the LB controller (e.g., Ananta, HAProxy), which in turn
programs the MUXes with the new weights. The MUXes may implement the
weights using WRR (weighted round robin). As shown in Fig.\ref{fig:overview:arch}, existing LB controllers are
\textit{not} on the critical path of the traffic\cite{ananta:sigcomm13,
  maglev:nsdi16}. Likewise, \name controller is not on the
critical path. \name does not result in any degradation in performance when MUXes handle packets.


\textbf{Discussion on heterogeneous requests and URL at KLM:} DIPs can handle heterogeneous requests that take  different paths (e.g., through cache or disk or other microservices). If DIP is overloaded, it will be reflected on the requests sent by the KLMs using the URLs set by admins. It may happen components \textit{other than DIPs} become bottleneck that inflates latency. Even then, \name will set the weights to optimize for latency. We want admins to set URL to \textit{most common type of requests} so that \name can capture performance more broadly. Currently, \name only supports one URL. We set it to future work to allow multiple URLs gauging performance of different paths.

\textbf{Discussion on why use weights and application latency:} An alternate approach could simply to monitor CPU usage on DIPs and steer traffic away from hot DIPs. However, it will fall short as: (a) it either requires agents on the DIPs or access to client subscription -- both raising privacy concerns and are non-goals for \name. (b) Application latency is not a function of just CPU utilization.  Application latency also depends on many factors across the stack such as CPU cache at hardware\cite{slomo:sigcomm20}, throttling at hypervisor as well as application related factors (such as queues). It is not trivial to capture  multitude resource counters and translate to application performance\cite{slomo:sigcomm20}. Instead, measuring application latency closely captures the service performance, and eliminates the need to capture and translate multitude signals from hardware to software. (c) We note that application performance depends on \textit{traffic going into DIPs} and different amounts of traffic could go to the same DIP for the same weights (as total traffic changes) resulting in different performance. We address such dynamics in \S\ref{sec:dynamics}.

\begin{table}[t]
\centering
\caption{Notations used in the ILP.}
\label{tab:algo:notation}
{
\small
\begin{tabular} {|c|c|}
\hline
\textbf{Notation} & \textbf{Explanation}\\
\hline
\hline
\multicolumn{2}{|c|}{Input} \\
\hline
$D$ & Set of DIPs \\
\hline
$w$ & Weight between [0,1]\\
\hline
$W_{d}$ & Set of weights for d-th DIP\\
\hline
$l_{d,w}$ & Latency on d-th DIP for w-th weight\\
\hline
$y_{max},$  & Max. and min. weights \\
$y_{min}$ & across DIPs\\
\hline
\hline 
(output) & Set if w-th weight is assigned \\
$X_{d,w}$ &  to d-th DIP\\
\hline
\end{tabular}
}
\end{table}

\subsection{Framing Load Balancing as a Knapsack Problem to Optimize Service Latency}
\label{sec:ilp}


Rather than a ``typical''
LB approach of spreading load evenly, we frame LB as a Knapsack problem\cite{knapsack:book} to pack the load as per DIP capacities to \textit{minimize overall service latency} (measured by KLMs above). To do so, we present an Integer Linear Program (ILP) that calculates weights to assign to DIPs to optimize for service response latency. The ILP  is shown in Fig.\ref{fig:algo:lp}. The
notations are listed in Table \ref{tab:algo:notation}. There is a
distinct ILP that assigns weights for each VIP.

\begin{figure}[t]
{\small
\fbox{
  \begin{minipage}{0.45\textwidth}
    \textbf{ILP Variable:} $X_{d,w}$\\ 
    \textbf{Objective:} Minimize $\displaystyle\sum_{d \in D}\displaystyle\sum_{w \in W_{d}} X_{d,w} \cdot l_{d,w}$\\
    \textbf{Constraints:}\\
    Only one weight for to each DIP: $\forall d \in D, \displaystyle\sum_{w \in W_{d}} X_{d,w} = 1$ \hfill (a)\\
    Total weight is 1: $ \displaystyle\sum_{d \in D}\displaystyle\sum_{w \in W_{d}} X_{d,w} \cdot w = 1$\hfill (b)\\
    Allowed imbalance: $y_{max} - y_{min} \leq \theta$ \hfill (c)
    $\forall d \in D, y_{max} \geq \displaystyle\sum_{w \in W_{d}} X_{d,w} \cdot w,  y_{min} \leq \displaystyle\sum_{w \in W_{d}} X_{d,w} \cdot w$ \hfill (d) 
  \end{minipage}
}
}
\caption{ILP formulation.}
\vspace{-0.15in}
\protect\label{fig:algo:lp}
\end{figure}

\label{sec:ilp:algo}


Rather than letting DIP weights vary arbitrarily in the interval $[0,1]$, \name uses a fixed set of possible weights that may be assigned to any DIP. The ILP variable $X_{d,w}$ (boolean) indicates if the $w$-th weight is
assigned to the $d$-th DIP. The objective is to minimize the total mean
latency (Fig.\ref{fig:algo:lp})\footnote{The objective can be easily changed to other objectives such as minimize max. latency or minimize sum of max. latency.}. Using a discrete set of weights (\eg $W_d = \{0.1, 0.2, ... 0.9,
1.0\}$) allows us to provide $l_{d,w}$,
the response latency at the $d$-th DIP if it is assigned the $w$-th
weight, as a numerical parameter to the ILP. The values for $l_{d,w}$ are computed before running the ILP (\S\ref{sec:design:curve}).

There are four constraints: (a) We assign only one weight to each DIP,
(b) the sum of total weight assigned across all DIPs is 1. (c) the
imbalance is restricted to $\theta$ (we do not aim to load balance
equally), (d) we specify $y_{max}$ and $y_{min}$ as the maximum and
minimum weights across all DIPs.

Knapsack problems in general are NP-complete\cite{knapsack:np}. We choose ILP approach as off-the-shelf ILP solvers optimize for a given objective, and have long been used for resource allocation problems\cite{ilp:1960, ilp:book}. However, in this paper, we show the challenges in using such ILPs for LB and our design choices to overcome those challenges. 

\subsection{Technical Challenges}
\label{sec:ilp:challenges}

\begin{figure}[t]
\centering
\includegraphics[width=0.44\textwidth]{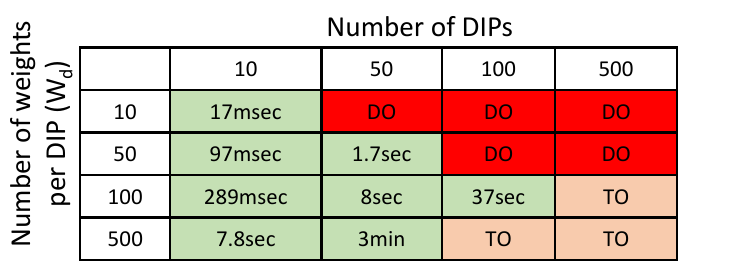}
\caption{ILP performance for varying \#DIPs and \#weights per DIP. DO and TO indicate DIP Overload and Timeout (ILP could not finish in 20 mins). }
\label{fig:ilp:runtime}
\end{figure}

There are three key challenges in implementing load
balancing using the ILP formulation in \S\ref{sec:ilp:algo}.


The first challenge is collecting the $l_{d,w}$, \ie the latency for
$d$-th DIP for $w$-th weight. As we do not know the latencies (performance)
beforehand, a strawman approach could be to measure latency uniformly
for $w$ in [0,1]. However, such an approach would require latency
measurements at large numbers of weights. Imagine two DIPs with
capacity 1$\times$ and 19$\times$ with roughly equal latency when not
loaded. The optimal weights for such a case are 0.05 and 0.95. Simply
measuring latencies at 11 points in range [0,1] (0,0.1,0.2, ..., 1)
will not yield an optimal split.  Worse, with these $W_d$, the ILP can
only calculate weights where at least one DIP is overloaded. Thus, we
need latency measured at weights with finer resolution. However,
increasing the resolution also increases the number of latency
measurements required per DIP. Further, since DIPs can have different
performance, we might not be able to reuse the measurements from one DIP to
another. For 100s of DIPs and measurements at 100s of weights/DIP
can make this approach prohibitively expensive.



The second challenge arises in measuring latency at even a single
weight. When we set the weight for a DIP, we cannot measure the
latency impact of this weight change right away, as we will need to
drain the existing connections for new weights to take effect
(\S\ref{sec:design:fct}). Coupled with above challenge of large number of measurements, it could take tens of
minutes to an hour just to collect all the necessary measurements. \ngs{$\leftarrow$ Do we actually solve this problem, and how? Even \name will incur drain time}


The third challenge is the computational feasibility of the ILP even
when all the measurements $l_{d,w}$ are available.
We run an experiment when all DIPs have the same
performance. Fig.\ref{fig:ilp:runtime} shows the time to compute the
weight assignment for varying number of DIPs and
weights per DIP ($W_d$). We speed up ILP as detailed in \S\ref{sec:impl}.
The weights are chosen uniformly between [0,1]. Even for a single VIP
of 500 DIPs, the ILP results in DIP overload (DO) (load for at least
one DIP is above its capacity) or timeout (TO) (ILP takes 20+ mins). Timeouts are especially worrisome as they affect the system's
responsiveness to failures and traffic dynamics
(\S\ref{sec:dynamics}). 

\section{Detailed Design of \name}
\label{sec:detailed-design}
\label{sec:design}

\subsection{Key Algorithms}
\label{sec:overview:key}

\name includes five algorithmic components to address the challenges
in \S\ref{sec:ilp:challenges}.

\textbf{C1. Curve fitting to reduce the number of latency
  measurements:} We observed that we do not need latency
measurements for many weights. We can perform just a small number of
measurements and use polynomial regression to build the curve and
estimate latency at other weights. \name includes a measurement phase where it builds a {\em weight-latency curve (\S\ref{sec:design:curve}).}

\textbf{C2. Adaptive weight setting based on prior latency
  measurements:} Complimentary to C1, our second algorithm adaptively
determines the next weight to use to conduct a latency measurement,
given existing weight-to-latency measurements. In particular, the
algorithm focuses the active measurements on DIP weights that are
likely to be more useful for load balancing, avoiding weights that
result in too-low or too-high response latencies. The algorithm is
loosely inspired by TCP congestion control. Together, C1 and C2
obviate the need for hundreds of measurements per DIP
(\S\ref{sec:design:measure}). As shown in \S\ref{sec:testbed}, \name
works with fewer than 10 measurements/DIP.


\textbf{C3. Multi-step ILP computation:} We need to decide on the set of
weights to feed to the ILP ($W_d$). Rather than feeding a large set of
possible weights in one shot to the ILP, we feed the weights in
multiple steps, with increasing resolution on the
weights, while holding the size of $W_d$ constant. Our algorithm is
detailed in \S\ref{sec:design:runtime}.  For the weights
$W_d$ chosen as candidate inputs for the ILP, we obtain $l_{d,w}$
using the polynomial regression for the weight-latency curve.

\textbf{C4. Quick reaction to dynamics:} A weight-latency curve
learned at a fixed aggregate load at the LB does not generalize to DIP
latencies at a different aggregate load. Online services exhibit many
dynamics including traffic changes, failures, and capacity changes. We
design mechanisms to quickly react to dynamics
(\S\ref{sec:dynamics}) by shifting the learned weight-latency curve
over time.

\textbf{C5. Scheduling measurements:} Even when the weights desired
for latency measurements are available (C2), we cannot assign those
weights to DIPs in a single shot. The weights may not add up to 1, and
different DIPs may pose different priorities in how soon measurement
data is needed. We show how to \textit{schedule} latency measurements
(\S\ref{sec:design:schedule}).

\subsection{Curve Fitting}
\label{sec:design:curve}

We seek to get the values of $l_{d,w}$ with a small number of latency measurements. We do so by making the latency measurements for small number of weights and implement \textit{curve fitting} (x-axis = weight, y-axis = latency) using polynomial regression. This way we can estimate the latency for other values of weights where we did not make latency measurements directly. Fig.\ref{fig:design:weight} shows the CPU utilization and latency as we vary the weights (traffic). It can be seen that  the latency increase is minimal at low weights as there is available CPU to handle the requests. We see higher increase in latency as weights increase as the CPU utilization hits 60\%. For higher weights when the CPU utilization hits 95\% or higher, we also observe packet drops (not shown). Such a relationship between latency and weight helps us do curve fitting using polynomial regression of degree two (\S\ref{sec:testbed}).

\subsection{Adaptive Weight Setting for Latency Measurement}
\label{sec:design:measure}

\begin{algorithm}[t]
\caption{Algorithm to calculate weights for latency measurement }
\label{alg:design:weight}
\begin{flushleft}
        \textbf{INPUT:} $l_{0}, l_{w}, w_{now}, w_{prev}, w^{prev}_{max}$ \\
        \textbf{OUTPUT:} $w_{next}, w_{max}, isExplorationDone$
\end{flushleft}
\begin{algorithmic}[1]
\If{$w_{now} - w_{prev} \leq D$}
    \State $isExplorationDone \gets 1$;
    return
\EndIf
\If{!(packet drop)}
    \State $w_{max} = max(w^{prev}_{max}, w_{now})$
    \State $w_{next} \gets w_{now} + w_{now} \cdot \alpha \cdot \frac{l_{0}}{l_{w}} $ \Comment{Run phase}
\Else
    \State $w_{next} \gets \frac{w_{now} + w_{prev}}{2}$ \Comment{Backtrack phase}
\EndIf
\end{algorithmic}
\end{algorithm}


The goals of the measurement phase are twofold: (a) identify a small
number of weights to perform latency measurements, so that the
measurements finish quickly and yet provide good curve-fitting, (b)
get a rough estimate for the capacity of the DIP in terms of the
weight (to calculate $W_d$). Recall that we do not know the capacities
of the DIPs beforehand.

Our algorithm for weight selection is inspired by TCP congestion
control and has two phases: (a) run, or (b) backtrack, depending on
the measured latency increase and packet drop. Algorithm
\ref{alg:design:weight} shows the algorithm to calculate weight for
each DIP in each iteration in the measurement phase.


The input to the algorithm includes $w_{now}$ and $w_{prev}$, the
weights for current and previous iterations. $w_{max}$ indicates the
maximum weight observed so far without packet drop. $w_{max}^{prev}$ is max. weight till last iteration. The input 
includes $l_{0}$ and $l_{w}$, the latencies when the weights are 0
and current weight. We measure $l_{0}$ when the DIP
is newly added by setting its weight to 0. The output includes: (a)
$w_{next}$ the weight whose latency should be measured in the next
measurement, (b) $isExplorationDone$, a boolean indicating if the
computation to get the weight-latency curve for this DIP is over, and it
is ready for the ILP to assign the weights for this DIP, (c) $w_{max}$.

As noted on line 1-2, if the difference between $w_{now}$ and
$w_{prev}$ is small (D = 5\% of $w_{now}$), we set the $isExplorationDone$ flag. Next, when
there is no packet drop, it indicates that there is still some
capacity remaining and we can increase the weight. We update the
$w_{max}$ (line 5) and increase the weight \textit{proportional to
  latency} (line 6). When $l_w$ is comparable to $l_{0}$, it indicates
there is more capacity left, and we can have bigger increase in
$w_{next}$. When $l_w$ is considerably higher than $l_{0}$, it
indicates we are reaching capacity, and we slow down the increase in
$w_{next}$. $\alpha$ indicates the pace of increase (set to 1 in
\name). When there is a packet drop (we have reached capacity), we
reduce the $w_{next}$ to the average of $w_{now}$ and $w_{prev}$ (line
8) and continue the search.  To improve exploration time and reduce
packet loss, we assume ``packet drop'' is $true$ on lines 4 and 7 when
$l_{w}$ is 5$\times$ $l_{0}$ based on the observation that latencies
are 5$\times$ $l_0$ or higher when the CPU is $\sim$100\% utilized.

We build the curve quickly where the total traffic change during curve building (few minutes) is usually small\cite{insidefb:sigcomm15}.

\subsection{Multi-Step ILP Computation}
\label{sec:design:runtime}

In steady state, once the weight-latency curve is available, we need to decide weights
in $W_{d}$. Once we decide weights, polynomial regression quickly
returns corresponding $l_{d,w}$ for the ILP. As shown in
Fig.\ref{fig:ilp:runtime}, the ILP running time increases rapidly with
the number of weights. Instead of running the ILP in one
step, we run in two steps, while providing a small number of weights
in each step. E.g., instead of running ILP with 100 weights for
every DIP, in the first step, we only provide 10 values uniformly in
$[0, w_{max}]$ (note, not [0, 1]). This provides a coarse estimate for
the weights without packet drop. In the second step, we calculate the
weights more precisely. If $w_d$ is the weight chosen by ILP for the
$d$-th DIP in first step, we provide 10 values uniformly between $w_d
- \delta$ to $w_d + \delta$ ($\delta$ = 10\% of $w_{max}$). We do
multi-step iteration only when the \#DIPs $\geq$100. Otherwise,
we do first step only. We program the LB dataplane only after the completion of both steps.

\subsection{Handling Service Dynamics}
\label{sec:dynamics}

We present mechanisms to address the drift in weight-latency curves
over time. 

\ngs{One thing that is a bit unclear in this section is that we assume that the load is fixed and nothing is changing when the first weight-latency curve is being built. Should we state that explicitly?}



\textbf{Addressing change in traffic:} When total traffic increases,
the traffic volume going to the DIPs will increase for the same
weights. Thus, we will observe higher latency for the same
weights. We detect the traffic change when we see latency increase for
most/all DIPs even when the weights are unchanged. We then ``shift'' the
weight-latency curve ``to the left'' as follows: let's say the latency
was 5 msec at weight 0.5 ($w_{1}$). With increased traffic, now the
latency has increased to 7 msec for the same weight. To calculate the
new weight, we multiply the existing weights by $\delta$. Let's say
the weight ($w_{2}$) for latency of 7 msec was 0.625. We calculate $\delta=\frac{w_{1}}{w_{2}}$. We multiply all the weights
with $\delta$. Similarly, when the traffic has reduced, we increase
the weights for the same latency values using above mechanism. 
\textbf{Addressing changes in capacity:} The capacity of the DIPs can
change dynamically (\eg due to a change in co-located VMs). We detect capacity change for a DIP if observed latency differs from the
estimated latency by more than a threshold (set to +/- 20\% of
$l_{d,w}$). We adjust the weights as detailed above. E.g., if the latency has changed from 5 msec to 7 msec, we use the
$w_1$ and $w_2$ from above mechanism.

\textbf{Addressing DIP failures:} \name detects DIP failures when we fail to get successful responses for KLM
probes. In such cases, we simply rerun ILP without that DIP.


\textbf{Refresh map:} We periodically refresh the weight-latency curve. At any point, we limit the fraction of DIPs under refresh to 5\% of the total capacity. To refresh, we simply measure
the latency for the weights as calculated in
\S\ref{sec:design:measure}.

We recalculate the weights for a VIP using new weight-latency
curve as the curve is updated for any of its DIPs.

\subsection{Scheduling Measurements}
\label{sec:design:schedule}
In \name, the weights for the DIPs with $isExplorationDone$ unset are
calculated using algorithm from
\S\ref{sec:design:measure}. However, we may not be able to measure latency
for the calculated weights right away, as the sum of all DIP weights
for a VIP needs to be 1. E.g., $w_{next}$ for
2 DIPs can be 0.7 each in one iteration. In such scenarios, we need to
\textit{schedule} the DIP weight in multiple rounds.

We classify DIPs with new weights
(with $isExplorationDone$ unset) to be scheduled into 3 priority
classes: (a) weights for over-utilized DIPs (DIPs with large latency), (b) weights for remaining DIPs, (c) weights during
refresh. Within a class, we use FIFO. To schedule the DIPs that have
new weights assigned, we use a simple greedy algorithm where we
arrange all DIPs according to their priority. We hop over the list of
DIPs until either: (1) the weight of DIPs scheduled is 1, or (2) we
exhaust all DIPs.

It can happen that the total weight for the scheduled DIPs is $\leq$ 1 (especially in case (2) above). If this occurs, we calculate
the remaining weights for DIPs as follows. Suppose the total weight
assigned by the scheduler is $w_{s}$. Hence, we need to assign $1 -
w_{s}$ using the remaining DIPs. For the DIPs where we have
$isExplorationDone$ flag set, we use the ILP with a modified
constraint (b) in Fig.\ref{fig:algo:lp} where the total weight is $1 -
w_{s}$. In cases where the ILP returns \textit{unsatisfiable} with
this constraint, we split the remaining weight (1 -- $w_{s}$) equally
across all remaining DIPs.


\subsection{Calculating Old Flow Completion Time}
\label{sec:design:fct}

Lastly, once we recalculate the weights for latency measurement, we
want to program those weights to the LB dataplane and measure the
latency. However, we may be unable to measure the latency right
away. There are delays due to (a) the LB controller
(Fig.\ref{fig:overview:arch}) taking time to program the dataplane,
and (b) only new connections adhering to the new weights once the
dataplane is reprogrammed (to preserve connection
affinity~\cite{ananta:sigcomm13, duet:sigcomm14,
  maglev:nsdi16}). In particular, the old connections directed to a
DIP due to the old weights continue to influence the DIP's latency
after the weight change, resulting in a clouded view of the impact of
the weight change.

\name's approach is to wait till old connections finish. However,
since the \name controller does not modify the MUXes or DIPs, it does
not know whether the old connections are completed. We calculate the
time between setting the weight and latency measurement (called
\textit{drain time}) by using some extreme settings: for a DIP, we
first set the weights high enough that the latency is high (time
$T_1$). Then we set the weight to 0 so that no new connections go to
this DIP. We continuously measure the latency until it reaches $l_{0}$
(time $T_2$). We calculate drain time as $T_2 - T_1$.  We measure the
drain time every 120 mins. 

\section{Implementation}
\label{sec:impl}

We briefly describe the implementation of the three key building
blocks of \name as shown in Fig.\ref{fig:overview:arch}: (a) KLM, (b)
latency store, (c) \name controller.

\textbf{KLM:} We have a VM image running for KLM  which
can be deployed in each VNET. The VNET admin sets the list of DIPs and
application URL in a file that KLM uses. KLM measures latency for every DIP in that VNET every 5 seconds using the
application URL provided by the admins, and reports average latency
over 100 requests (independent of DIP size). We do not consider higher percentiles (such as P90
or P95) because we observed high and variable latency at high
percentiles even when the load is not high. Also, we found that
average latency correlates better with load (Fig.\ref{fig:design:weight}).


\textbf{Latency store:} KLMs write the latency to latency store. We use Redis\cite{redis:web} for latency store as it provides in-memory caching and fault tolerance. The key is VIP and value is list of \texttt{<DIP,latency,time>} tuples. The latency store runs in the same datacenter as VNETs and \name controller.

\textbf{\name controller:} \name controller is the heart of \name. It
consists of modules to (a) calculate weights for latency measurements,
(b) scheduling the latency measurements, and (c) running ILP to
calculate optimal weights. All the modules (for a VIP) run on the same
VM. \S\ref{sec:eval:overhead} provides the
overheads for running the controller for multiple VIPs.

The ILP uses the open source ILP solver
COIN-OR\cite{coinor:web},  written in C++ with PuLP
bindings\cite{pulp:web}.
Together, \name uses 4K+ LOC using high-level languages. For multiple VIPs, we prioritize
ILP for VIPs with a change in the weight-latency curve for
some DIP.
The controller by default runs ILP for each VIP every
5 seconds.

\section{Evaluation}
\label{sec:testbed}

\begin{table}
\centering 
\caption{VM details used in evaluation.}
\label{tab:eval:setup}
\small
\begin{tabular}{|c|c|c|c|c|}
\hline
DIPs & DIP-1 to  & DIP-17 to  & DIP-25 to  & DIP-29,30\\
 & DIP-16 & DIP-24 & DIP-28 & \\
\hline
VM type & DS1v2 & DS2v2 & DS3v2 & F8sv2 \\
\hline
\#vCPUs & 1 & 2 & 4 & 8 \\
\hline
\#VMs & 16 & 8 & 4 & 2\\
\hline
\end{tabular}
\end{table}

We evaluate \name using testbed experiments and simulations for large
number of DIPs. Our experiments show that (a) \name builds the
weight-latency curve fast and requires few points to build the curve;
(b) \name is able to compute weights optimally using the ILP to
minimize the overall latency \textit{without any hints} about the
capacity or performance. This is particularly useful as it allows users to add DIPs
of any capacity; (c) \name
substantially improves the latency compared to (weighted) LB policies -- many used in production; (d) \name handles the dynamics such as
failures, traffic and capacity changes well; (e) \name can work with
other LBs with and without interface to program weights; (f) \name
incurs very small overhead in terms of extra resources and costs.

\textbf{Setup:} Our setup consists of 41 VMs running in Azure central
US datacenter. There are 30 DIPs with different capacities (Table \ref{tab:eval:setup}), 1 8-core VM running HAProxy and 8 VMs as clients. Lastly, we have 1 VM each for \name controller
and KLM. We use Redis cloud offering\cite{azureredis:web}. The DIPs
run web server doing cache intensive calculation for client requests and clients: (a) send the requests to DIPs through
HAProxy, (b) measure the end-to-end latency.  All VMs run Ubuntu
20.04.  As shown in Table \ref{tab:eval:setup}, we use DIPs of 4 different
types. We specifically use VMs across different series (DS and F). The
F-series VMs\cite{fazure:web} are supposed to be faster than DS-series VMs\cite{dsazure:web} by up to
2$\times$. However, our measurements found that F-series VM is
15-20\% faster than corresponding DS-series VM. This also
highlights that it is not trivial to calculate the weights just by
considering \#cores and clock speed. A system like \name can
help automatically calculate and adjust weights to optimize performance.
We set the traffic to 70\% of total capacity. We set $\theta = \infty$
in Fig.\ref{fig:algo:lp} to not put any restriction and optimize for
latency.

\subsection{Weight Assignment in \name}
\label{sec:testbed:curve}

\begin{figure*}
\centering
    \begin{minipage}{.3\textwidth}
        \centering
        \includegraphics[width=1\textwidth, page=1]{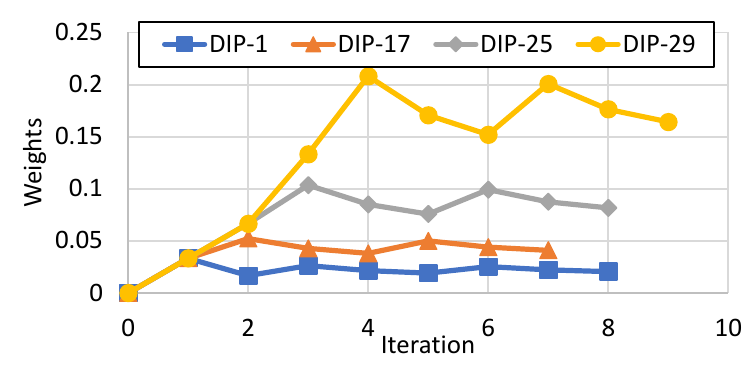}
        \vspace{-0.2in}
        \caption{Weights used for latency measurements.}
        \label{fig:eval:weights}
    \end{minipage}%
    \hspace{0.2cm}
    \begin{minipage}{0.3\textwidth}
        \centering
        \includegraphics[width=1\textwidth, page=1]{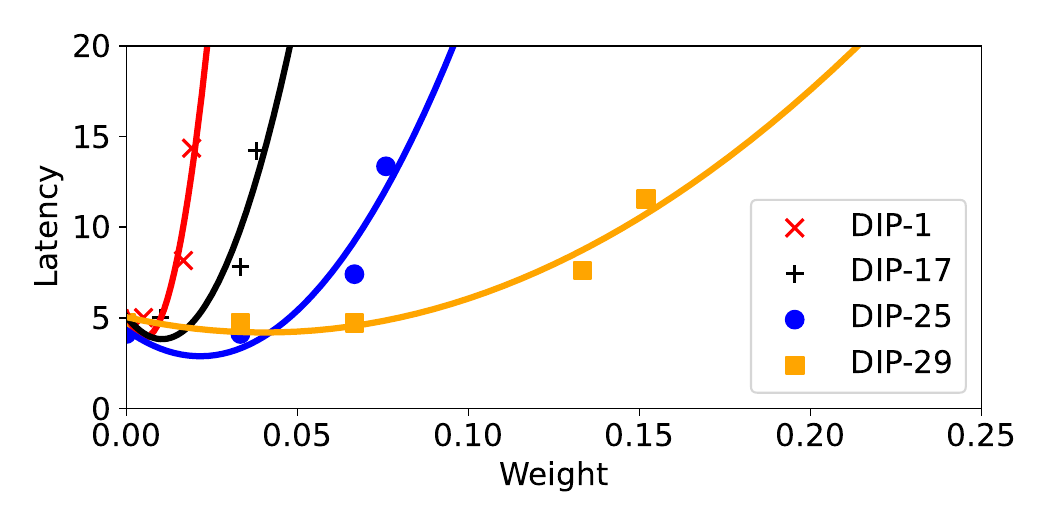}
        \vspace{-0.2in}
        \caption{Curve fitting using polynomial regression.}
        \label{fig:eval:poly}
    \end{minipage}%
    \hspace{0.2cm}
    \begin{minipage}{0.3\textwidth}
        \centering
        \includegraphics[width=1\textwidth,page=2]{figs/eval-weight-hopping.pdf}
        \vspace{-0.2in}
        \caption{Weights calculated by ILP.}
        \label{fig:eval:ilp}
    \end{minipage}%
\vspace{-0.2in}
\end{figure*}

\textbf{Calculating weights for latency measurements:}
We start our evaluation by calculating the weight-latency curve for all the 30 DIPs. We first start by measuring the latency when weight is 0 and equal (0.033 in this experiment). We then adjust the weights using algorithm \ref{alg:design:weight}.


Fig.\ref{fig:eval:weights} shows the different weights calculated for
4 different types of DIPs. We randomly chose one DIP from each type
(Table \ref{tab:eval:setup}). We observe: (a)
it took 8-10 iterations to build weight-latency curve for all
DIPs. Here, we only show the weight calculated by the algorithm
\ref{alg:design:weight}, and not by the scheduler
(\S\ref{sec:design:schedule}). Thus, the total weight per
iteration is not 1. (b) Next, for each iteration, the weights are
scheduled in multiple rounds. The scheduler took 1.7 rounds for each
iteration (on average). Each round was 10 seconds. Therefore, the
entire experiment finished in less than 3 mins. (c) the weights
calculated for measurements vary across DIP types. Also, the $w_{max}$
for the 4 DIPs calculated are 0.02, 0.04, 0.085, 0.165. Note, $w_{max}$ corresponds to max. weight \textit{without} packet drop. Thus, the values are lower than the peak weight calculated. 

%

\textbf{Weight-latency curve using polynomial regression:} After we measure the latencies for different weights, we use polynomial regression to fit the curve so that we can estimate the latencies for weights not used in measurements. For polynomial regression, we only use the data for which there was no packet drop. As a result, there are only 4 points for DIP-1 to DIP-28 and 5 points for DIP-29,30. Fig.\ref{fig:eval:poly} shows the weight-latency curve for the 4 DIPs of 4 different types. The points show the actual measurements, while the lines show the curve calculated by polynomial regression. It can be seen that the regression fits the curve well even when it has only a small number of points.  As we increase the weights, we expect the latency to go up. However, as can be seen in the Fig.\ref{fig:eval:poly}, the latency may not increase monotonically using  regression. To address this limitation, we change the regression output to increase monotonically by setting the latency for a given weight as max. of its latency (by regression) and latency at the previous weight (not shown in figure).

\textbf{ILP calculation:} Next, we calculate the weights using ILP
described in \S\ref{sec:ilp}. Fig.\ref{fig:eval:ilp} shows the weights
calculated for the 15 DIPs (we select 50\% DIPs from each type). The weights are in ratio 1:2:3.9:9.7. ILP
assigned more weight to the VM with more capacity (8 core). ILP
assigned weight of 0.135 to DIP-29 even though it has 12.5\% of total
capacity, and it assigned combined weight of 0.225 to DIP-1 to DIP-16
that together had 25\% of total capacity. This is because \name
optimizes for \textit{total latency}. 

\subsection{Comparing with Other LB Policies}
\label{sec:eval:rr}

\begin{figure}[t]
\centering
\subfigure[CPU utilization]
{
\includegraphics[width = 0.35\textwidth, page=2]{figs/eval-15-node-2.pdf}
\label{fig:eval:tbcpu}
}
\\[-0.1in]
\subfigure[Latency]
{
\includegraphics[width = 0.35\textwidth, page=1]{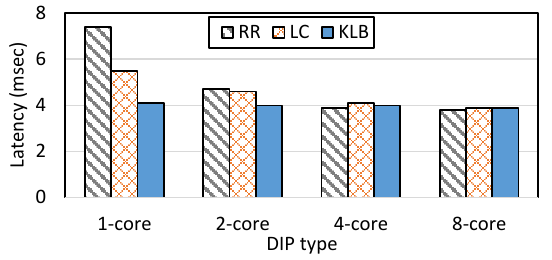}
\label{fig:eval:tblatency}
}
\vspace{-0.1in}
\caption{Average CPU and latency using 30 DIPs testbed.}
\protect\label{fig:eval:tb}
\vspace{-0.1in}
\end{figure}

\begin{figure}[t]
\centering
\includegraphics[width=0.35\textwidth, page=1]{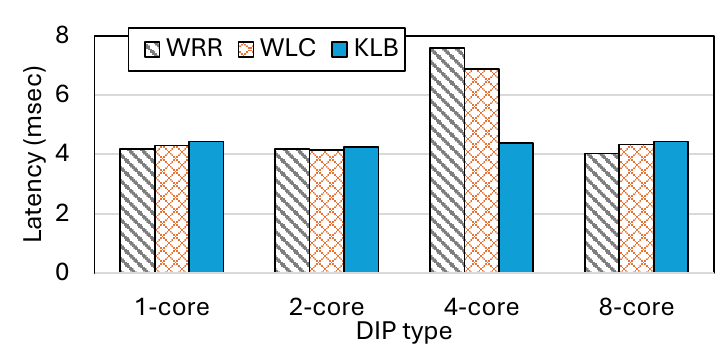}
\caption{Average latency using weights for 30-DIP cluster. 8-core VM is F-type. KLB denotes \name.}
\label{fig:eval:tbcpu:weights}
\end{figure}

\begin{figure}[t]
\centering
\includegraphics[width=0.43\textwidth, page=1]{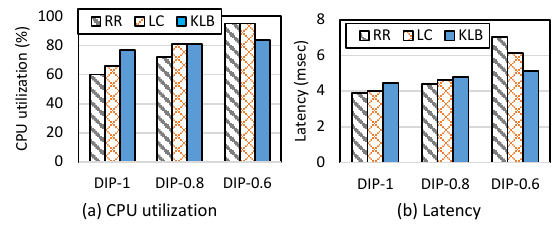}
\vspace{-0.1in}
\caption{Average CPU and latency using 3 DIPs of capacity 1$\times$, 0.8$\times$ and 0.6$\times$. KLB denotes \name.}
\label{fig:eval:tb3}
\end{figure}

\begin{table}[t]
\centering 
\caption{Max. gains in latency in \name compared to other LB policies with 30 nodes (\S\ref{sec:eval:rr}). RD = Random, P2 = Power-of-2.}
\label{tab:eval:allpolicies}
\small
\begin{tabular}{|c|c|c|c|c||c|}
\hline
 & RR & LC & RD & P2 & Azure\\
 \hline
 unweighted & 45\% & 23\% & 42\% & 24\% & 41\%\\
 \hline
 weighted & 42\% & 36\% & 41\% & NA & NA \\
\hline
\end{tabular}
\end{table}

We now show the improvements in \name compared to Least Connection (LC) and Round Robin (RR), Random (RD) and power-of-two (P2) policies -- both unweighted and weighted versions. RD selects DIP uniformly at random; P2 randomly picks two DIPs and then selects one DIP for a connection that has smaller CPU utilization among the two DIPs. Note that these policies are supported by existing LBs running in production (e.g., HAProxy and NGINX support them). Additionally, we also show the improvements over LB used by Azure LB. The metrics of interest are end-to-end latency observed by clients and CPU utilization on the DIPs. We show the average for 100K requests.  We use two DIP-pools: (a) same DIP-pool of 30 DIPs as before. (b) smaller DIP-pool of 3 nodes from \S\ref{sec:back:limit1}. 

\textbf{30 node DIP-pool with no weights:} Fig.\ref{fig:eval:tbcpu} shows the CPU util. across all DIP types for RR, LC and \name (KLB). Unsurprisingly, as RR is not optimized for performance, it results in high CPU utilization for DIP-1 to  DIP-24.  Conversely, DIP-25 to DIP-30 show lower CPU utilization as they receive lesser traffic compared to their capacity. LC improves the CPU utilization compared to RR by sending fewer connections to overloaded DIPs but it still results in higher CPU utilization for DIP-1 to DIP-16. In contrast, \name  results in uniform CPU utilization across all DIPs by being performance-aware.

Fig.\ref{fig:eval:tblatency} shows the latency across the 4 types of DIPs. In RR, we observe higher latency for DIP-1 to DIP-24 as such DIPs are overloaded. In contrast, latency from DIP-25 to DIP-30 is low as such DIPs are underloaded. LC improves the latency but it still results in high latency due to overloaded DIP-1 to DIP-16. In \name, as the DIPs are not overloaded, we observe low (and uniform) latency across all DIP types.  This experiment shows that \name cuts the latency by up to 45\% for 79\% requests compared to RR, and up to 23\% for 68\% requests compared to LC.

Table \ref{tab:eval:allpolicies} shows the max. gain using KLB over other policies including Random (RD) and power-of-2 (P2).

\textbf{Improvement over Azure LB:} As mentioned previously, Azure L4LB only supports equal LB through hash over TCP/IP fields. It assumes equal weight for all DIPs, which is problematic as we have DIPs with different capacities (number of cores). As a result, due to equal splitting, requests going to the DIPs with lesser number of cores observe higher latencies. We found that \name improves the latency by up to 41\% for 71\% requests compared to Azure LB.

\textbf{30 node DIP-pool with weights:} In the previous experiment, we did not use the weights. However, service operators can set the weights using hardware properties. In this experiment, we set the weights for RR and LC policies (denoted as WRR and WLC) in proportion to the number of cores. Recall that \name does not require such information apriori. Fig.\ref{fig:eval:tbcpu:weights} shows the average latency across all DIP types for WRR, WLC and \name (KLB). We found that the throughput of 4-core DS-type VM did not scale linearly with number of cores (8-core F-type VM also did not scale linearly but in general had more capacity). As WRR is not performance aware, its traffic to 4-core DIPs was unabated causing the DIPs to be overloaded and caused high latency. WLC reduced the traffic to such DIPs to a small extend. In contrast, \name reduced the traffic to such DIPs that lowered  latency. Compared to WRR and WLC, \name reduced the latency by 42\% and 36.2\% on such DIPs. 

Table \ref{tab:eval:allpolicies} shows the max. gains in \name for latency over other (weighted) policies. Power-of-2 (P2) and Azure do not support weights.

\textbf{3 node DIP-pool with weights:} In the next experiment, we measure the latency and CPU utilization based on the DIP-pool from \S\ref{sec:back:limit1} running on 1-core VMs. However, we change the capacity to emulate noisy neighbors -- we use capacities of 1$\times$, 0.8$\times$ and 0.6$\times$. In this experiment, we use weighted RR and LC with weights set to number of cores (1:1:1). Note that, the impact of noisy neighbors could be dynamic, variable and unpredictable. As shown in Fig.\ref{fig:eval:tb3}, both (weighted) RR and LC fall short in load balancing the traffic as per capacities, and over-utilize DIP-0.6 (DIP with capacity 0.6$\times$) while there is capacity available on other DIPs. As a result, we also observed higher latency on DIP-0.6. 

In contrast, \name (KLB) substantially improves the load balancing. It reduces the CPU utilization on the DIP-0.6 while making use of the available CPU on DIP-1. As a result, we observed uniform CPU on all the three nodes. In doing so, KLB improved the latency for all the connections that RR and LC sent to DIP-0.6. Compared to RR and LC, KLB cut latency by up to 37\% and 29\%.

\subsection{Handling Dynamics}

\begin{figure*}
\centering
    \begin{minipage}{.3\textwidth}
        \centering
        \includegraphics[width=1\textwidth, page=3]{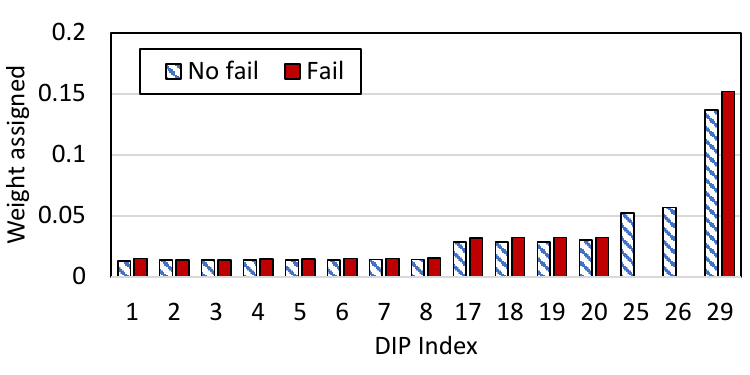}
        \vspace{-0.2in}
        \caption{Weight change due to failure of DIP-25,26.}
        \label{fig:eval:dyn:fail}
    \end{minipage}%
    \hspace{0.2cm}
    \begin{minipage}{0.3\textwidth}
        \centering
        \includegraphics[width=1\textwidth, page=5]{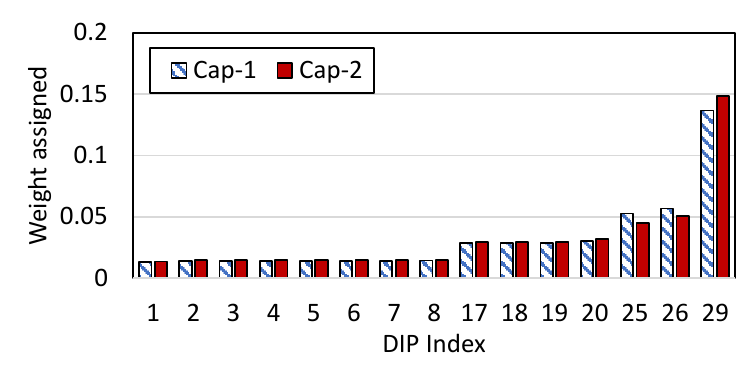}
        \vspace{-0.2in}
        \caption{Weights change as capacity changes for DIP-25 to -28. }
        \label{fig:eval:dyn:cap}
    \end{minipage}%
    \hspace{0.2cm}
    \begin{minipage}{0.3\textwidth}
        \centering
        \includegraphics[width=1\textwidth, page=4]{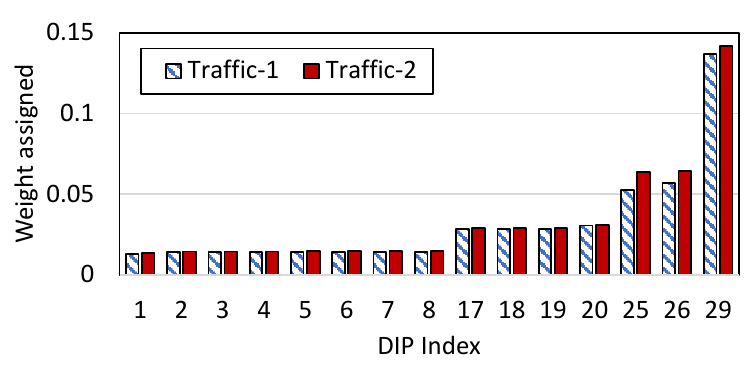}
        \vspace{-0.2in}
        \caption{Weight change due to traffic change.}
        \label{fig:eval:dyn:traffic}
    \end{minipage}%
    \vspace{-0.2in}
\end{figure*}

We now show the change in weights  as we address dynamics due to: (a) failures, (b) change in capacity and (c) change in traffic. We use the DIP-pool with 30 DIPs from Table \ref{tab:eval:setup}. 

\textbf{Addressing failures:} In this experiment, we fail DIP-25,26
while keeping the traffic unchanged. Fig.\ref{fig:eval:dyn:fail} shows
the weights before and after the failure. We observed that the weight
of the failed DIP \textit{was not} equally split among other
DIPs. DIP-1 to DIP-16, and DIP-17 to DIP-24 saw a cumulative increase
of 0.012 and 0.027 respectively in weights. Most of the weight of the
failed DIP was assigned to DIP-27 to DIP-30 (cumulative increase of
0.066), mainly because such DIPs with higher capacity had more room to absorb traffic with respect to the
increase in latency. None of the VMs were overloaded. The weights were
not equally split, as the ILP made latency-informed decisions.


\textbf{Addressing change in capacity:} Next, we bring back DIP-25,26 from last experiment. We reduce the capacity of DIP-25 to DIP-28 by co-running a process that consumes 1 core.  Total traffic is unchanged. This change in capacity was reflected in latency differences for the same weights, and we update the weight-latency curve for DIP-25 to DIP-28 (\S\ref{sec:dynamics}). Fig.\ref{fig:eval:dyn:cap} shows that \name is able to react to the capacity change and adjust the weights. Interestingly, instead of reducing the weights of DIP-25 to DIP-28 by 25\%, \name reduced the weights assigned to these DIPs by 15-17\%. The remaining weight was mostly assigned to DIP-29,30 and other DIPs saw a small increase. This is again due to better latencies on DIP-29,30 for the same weights.




\textbf{Addressing change in traffic:} In this experiment, we have all the DIPs at their original capacity, and we increase the traffic by 10\%. We detected traffic increase when all DIPs observe increase in latency for the same weights. In such cases, we shift the weight-latency curve to left. As shown in Fig.\ref{fig:eval:dyn:traffic}, it can be seen that DIP-25 to DIP-30  absorbed most of the extra traffic. This is again because such DIPs have more room to absorb traffic than other DIPs for the same latency increase. In all cases, \name load balanced the traffic as per capacities without overloading any DIP.


As we performed latency measurements every 5 seconds, the above dynamics were detected within 5 seconds. The ILP took roughly 120 msec to calculate the new weights. These experiments demonstrate that mechanisms for reactions to dynamics are quick and effective. 

\subsection{Comparing Against Agent-based Method}
\label{sec:eval:agent}

\name does not require any agents due to privacy reasons; we compare \name against agent-based method where we have an agent on each DIP to measure its CPU utilization and adjust the load to get uniform CPU utilization. Note that, as mention in \S\ref{sec:back} and \S\ref{sec:overview}, simply measuring CPU utilization falls short in optimizing for performance as performance depends on many factors including hardware contention (cache, memory bus etc.), throttling at hypervisor, application queues and more. Despite this, we consider such a baseline. We use 4 DIPs of same VM-type but reduce the capacity of one DIP to 75\%. We use an algorithm to compute weights from sec. 4.1 in \cite{cheetah:nsdi20}, which computes weights iteratively. Such an algorithm took 4 iterations to get weights for uniform CPU utilization, whereas \name just took one iteration using ILP.


\subsection{Load Balancing using Other LBs}
\vspace{-0.1cm}

\label{sec:eval:other}

\begin{table}[t]
\begin{minipage}[b]{0.45\textwidth}
\centering 
\caption{Fraction of requests received by 3 DIPs using Nginx and Azure traffic manager (TM).}
\label{tab:eval:otherlb}
\begin{tabular}{|c|c|c|c|}
\hline
LB & DIP-1 & DIP-2 & DIP-3 \\
\hline
Nginx & 20\% & 30\% & 50\%\\
\hline
Azure TM & 18\% & 34\% & 48\%\\
\hline
\end{tabular}
\end{minipage}
\hspace{0.5cm}
\begin{minipage}[b]{0.45\textwidth}
\centering 
\caption{ILP running time (msec) across different \#DIPs.}
\label{tab:eval:runtime}
\small
\begin{tabular}{|c|c|c|c|c|c|}
\hline
\#DIPs & 10 & 50 & 100 & 500 & 1000 \\
\hline
time (msec) & 20 & 194 & 645 & 5.8K & 21.1K \\ 
\hline
\end{tabular}
\end{minipage}
\vspace{-0.1in}
\end{table}

Now, we demonstrate that \name can work with: (a) Nginx\cite{nginx:web}: another widely used L4 LB that provides an interface to specify the weights. (b) Azure: LB does not provide an interface to specify weights. In (b), we use traffic manager (TM) to do load balancing using DNS\cite{azuretm:web} that resolves the IP based on the weights of the DIPs. In this experiment, we use 3 DIPs behind the above LBs with weights DIP-1 = 0.2, DIP-2 = 0.3, DIP-3 = 0.5.  Table \ref{tab:eval:otherlb} shows the fraction of requests received by individual DIPs (total requests = 10K). Nginx does LB as per the weights specified and can work with \name like HAProxy. For Azure TM, it roughly splits the DNS requests in the weights specified. However, we note that such a load balancing depends on the DNS cache timeout, and clients can see delay in adhering to new weights. This experiment shows that \name can work with other LBs, also using DNS when LBs dont provide a native interface to specify weights. 

\subsection{Simulation with Large Number of DIPs}
\label{sec:eval:sim}

Now we turn to ILP running time and accuracy of multi-step ILP. We use
simulations for large number of DIPs. For each DIP, we keep the
capacity the same and use the latency-weight curve for the F-series VM
from the previous section, and the traffic set to 80\% of the total
capacity. 

\textbf{ILP running time:} We measure the ILP
running time (shown in Table \ref{tab:eval:runtime}) as we vary the number of DIPs from 10 to 1000. We feed 10
points uniformly between 0 and $w_{max}$. It can be seen that ILP runs quite fast: when
the number of DIPs is smaller than 100, the ILP completes in
645msec. When the number of DIPs is 1000, the ILP finishes in 21sec. 

\begin{table}[t]
\begin{minipage}[b]{0.95\linewidth}
\centering
\caption{Accuracy and running time in multi-step ILP. For 10 \#points, we run ILP twice.}
\label{tab:eval:multistep}
\begin{tabular}{|c|c|c|}
\hline
\#points & running time & accuracy \\
\hline
100 & 36.8sec & 100\% \\
\hline
10 & 0.65sec x2 & 99.9\% \\
\hline
\end{tabular}
\end{minipage}
\hspace{0.1cm}
\begin{minipage}[b]{0.95\linewidth}
\centering
\caption{Workload details showing \#VIPs for different \#DIPs/VIP for a 60K DIP datacenter.}
\label{tab:eval:workload}
\begin{tabular}{|c|c|c|c|c|c|c|}
\hline
\#DIPs/VIP & 5 & 10 & 50 & 100 & 500 & 1000 \\
\hline
\#VIPs & 2000 & 1000 & 200 & 100 & 20 & 10 \\
\hline
\end{tabular}
\end{minipage}
\end{table}

\textbf{Multi-step ILP:}  We now compare the accuracy and running time
by using multi-step ILP as described in
\S\ref{sec:design:runtime}. There are 100 DIPs. First, we feed 100
weights uniformly between 0 and $w_{max}$. Next, we feed in 10 weights
in two steps. We choose 10 as a sweet spot between speed and available weight options. 
We found that multi-step ILP reduces the
run-time by 28.3$\times$ while sacrificing only 0.1\% accuracy (Table \ref{tab:eval:multistep}).

\subsection{Overheads at Large Number of DIPs}
\label{sec:eval:overhead}

\textbf{KLM:} KLM sends the latency measurement probes (100 HTTP
requests) to individual DIPs every 5 seconds independent of the DIP size. We found the throughput
of KLM is 4500 requests/sec on DS1 v2 VM (1 core). The throughput
translates to supporting 225 DIPs per KLM VM. However, many VIPs may
have DIPs fewer than 225\cite{duet:sigcomm14} but would still require
a separate instance of KLM (due to VNET boundaries). As the \#DIPs/VIP
and \#VIPs is not publicly available, we use the workload detailed in
Table \ref{tab:eval:workload} consisting of many mice and a few
elephant VIPs, derived from \cite{duet:sigcomm14}. Even after
considering separate KLM for VIPs $\leq$225 DIPs, we would
need 3410 KLM cores for the 60K DIPs from table
\ref{tab:eval:workload}. Assuming DIPs run on D8a type (8 cores; \$280/month), and KLM runs on DS1 type (1 core; \$41/month), the
overhead in terms of \#cores and cost is just 0.71\% and
0.83\% respectively. KLM can run on spot VMs reducing costs by 2.6$\times$\cite{spotvm:web}.

\textbf{Latency store:} we use Azure Redis cache in the same DC with \textit{Premium} option\cite{redisprice:web} for good performance.  Each Redis \textit{get} operation takes 0.3-4 msec (using persistent connections). For 100K DIPs with 10 latency points per DIP, the total data easily fits within 6GB costing just \$6/day (with discount for 3 years subscription), which is a very minuscule overhead.  

\textbf{\name controller:} the controller runs: (a) polynomial regression, (b) ILP. For regression, it takes on average 1 msec/DIP on single core. For doing regression for 60K DIPs, it would take 60 cores at the controller. With 8 cores/DIP, the overhead (as number of cores) is just 0.01\%. Next, the total running time to run ILP for the workload detailed in Table \ref{tab:eval:workload} is 851 seconds on a single 8-core VM. To run the ILP per VIP every 5 seconds, we would need 193 such VMs (accounting for VIPs that take $\geq$ 5 seconds to finish ILP). With all DIPs and controller running on 8 core VMs, the overhead (in terms of number of cores and cost) is just 0.32\%.

In summary, \name imposes very small overhead (extra cores
and costs) while providing substantial benefits in packing the service load according to dynamic DIP capacities.

\section{Related Work}
\label{sec:related}

To the best of our knowledge, \name is a novel approach to build performance optimized L4 LBs. There has been much work on performance variability and LBs:

\textbf{Performance variability and prediction:} Performance variation
has been acknowledged for many years. Google reported up to 20\%
capacity reduction due to noisy
neighbors\cite{googlelb:web}. Similarly, \cite{slomo:sigcomm20,
  resq:nsdi18, cpusteal:cloud15} report up to 40\% change in
capacity. \cite{dcstore:fast20, mt2:fast22} report up to 450\%
degradation in performance due to noisy neighbors in storage. Other works also shed light on performance variability (\cite{uta:nsdi20, cao:fast17, perf:lambda:wosc20, taming:osdi18, temp:sigmetrics12, flash:sigmetrics15, hardware:dsn17, money:socc12}). Recent
works have also focused on predicting the performance due to resource
contention (\eg Ernest\cite{ernest:nsdi16} for
analytic workloads on cloud, Paris\cite{paris:socc17} for choosing best VMs in cloud). However, they
use offline profiling of workloads. In contrast, \name is completely
online and does not require any profiling
apriori. \cite{slomo:sigcomm20, resq:nsdi18, predictnf:nsdi12} provide
performance variation and prediction for NFV, but rely on access to hardware counters (e.g., cache misses) that
unfortunately are not available to tenants (or to \name) in the
cloud. That said, even if such counters become available, adjusting
weights based on the counters remains challenging.

\textbf{LB designs:} Recent works on LB focus on cost, availability, scalability. Ananta\cite{ananta:sigcomm13} and Maglev\cite{maglev:nsdi16} are running in production. \cite{duet:sigcomm14, silkroad:sigcomm17, tiara:nsdi22} use hardware to save costs. \cite{cheetah:nsdi20, beamer:nsdi18} focus on improving the resiliency of LB. \name is complementary to these designs and adjusts the weights based on capacity. \cite{nginx:lb20, hlb:ton22, shobhana:hotnets22} rely on DIP counters or changes to network/MUXes for load balancing.

\textbf{Latency as a congestion signal:} Prior works have also used latency as a signal for congestion control including Timely\cite{timely:sigcomm15}, Swift\cite{swift:sigcomm20} and BBR\cite{bbr:2017}. \name uses the latency as congestion signal for load packing.  

\textbf{Capacity based packing:} Prior works have looked at capacity based packing but in different context. E.g., multi-resource cluster scheduling \cite{tetris:sigcomm14}, VM/container placement \cite{vmpacking:iwqos13, vmpacking:spaa14, draps:IPCCC}, CDN routing \cite{fastroute:nsdi15}, replica assignment \cite{pegasus:osdi20, c3:nsdi15}. Such works use intrinsic resource usage, feedback from the servers and/or assistance from clients. E.g., \cite{c3:nsdi15} requires servers to send queue sizes and clients to rank servers. \cite{prequal:nsdi24} also works with queues tied to the applications and requires changes to servers and LB. \cite{fastroute:nsdi15} requires changes to MUXes, DIPs to gather and dissipate load details and rerouting. \name neither requires such information nor control on DIPs, MUXes and clients. 

\section{Conclusion}
\label{sec:conc}
\vspace{-0.05in}
We present \name to empower other layer-4 LBs to balance load according to the
performance of backend instances (DIPs). \name is agent-less, fast and versatile -- suitable to use with a wide variety of LBs. \name builds
weight-latency curves, formulates an ILP problem to minimize overall latency, and proposes
techniques to expedite the ILP computation. Using prototype implementation and large-scale simulations, we show \name cuts latency by up to 45\%.

\bibliographystyle{abbrv}
\bibliography{references}

\end{document}